\documentclass[11pt]{article}

\usepackage[T1]{fontenc}
\usepackage[utf8]{inputenc}

\usepackage[margin=1in]{geometry}

\usepackage{amsmath,amssymb,amsfonts,amsthm}

\usepackage{graphicx}
\usepackage{booktabs}
\usepackage{multirow}
\usepackage{tabularx}
\usepackage{threeparttable}
\usepackage{float}

\usepackage{algorithm}
\usepackage{algpseudocode}

\usepackage[hidelinks]{hyperref}

\usepackage{authblk}

\usepackage[title]{appendix}
\usepackage[numbers,sort&compress]{natbib}

\usepackage{xurl}
\usepackage{microtype} 

\usepackage[hidelinks]{hyperref}

\usepackage{etoolbox}
\AtBeginEnvironment{thebibliography}{\sloppy\raggedright}
\theoremstyle{plain}

\theoremstyle{definition}

\theoremstyle{remark}

\newcommand{\zooNet}{\textsc{zooNet}}

\title{From Ecological Connectivity to Outbreak Risk: A Heterogeneous Graph Network for Epidemiological Reasoning under Sparse Spatiotemporal Data}
\date{} 

\author[1,2]{Haley Stone}
\author[1]{Jing Du}
\author[1]{Yang Yang}
\author[1]{Ashna Desai}
\author[2]{Rebecca Dawson}
\author[1]{Hao Xue}
\author[3]{David Heslop}
\author[4]{Matthew Scotch}
\author[5]{Andreas Z{\"u}fle}
\author[2]{C.\ Raina MacIntyre}
\author[1]{Flora Salim}

\affil[1]{Computer Science and Engineering, Faculty of Engineering, University of New South Wales, Kensington NSW 2052, Australia}
\affil[2]{The Kirby Institute, Faculty of Medicine \& Health, University of New South Wales, Kensington NSW 2052, Australia}
\affil[3]{School of Population Health, Faculty of Medicine \& Health, University of New South Wales, Kensington NSW 2052, Australia}
\affil[4]{College of Health Solutions, Arizona State University, Phoenix AZ 85004, United States}
\affil[5]{Department of Computer Science, Emory University, Atlanta GA 30322, United States}

\begin{document}
\maketitle

\begin{center}
\small
\textbf{Corresponding Authors:} \texttt{h.stone@unsw.edu.au \& flora.salim@unsw.edu.au)} \\
\end{center}

\vspace{0.75em}

\begin{abstract}
Estimating population-level prevalence and transmission dynamics of wildlife pathogens can be challenging, partly because surveillance data is sparse, detection-driven, and unevenly sequenced. Using highly pathogenic avian influenza A/H5 clade 2.3.4.4b as a case study, we develop \zooNet, a graph-based epidemiological framework that integrates mechanistic transmission simulation, metadata-driven genetic distance imputation, and spatiotemporal graph learning to reconstruct outbreak dynamics from incomplete observations.

Applied to wild bird surveillance data from the United States during 2022, \zooNet recovered coherent spatiotemporal structure despite intermittent detections, revealing sustained regional circulation across multiple migratory flyways. The framework consistently identified counties with ongoing transmission weeks to months before confirmed detections, including persistent activity in northeastern regions prior to documented re-emergence. These signals were detectable even in areas with sparse sequencing and irregular reporting.

These results show that explicitly representing ecological processes and inferred genomic connectivity within a unified graph structure allows persistence and spatial risk structure to be inferred from detection-driven wildlife surveillance data.
\end{abstract}


\section{Introduction}\label{sec1}
Zoonotic diseases constitute one of the most persistent and complex challenges in infectious disease ecology. The majority of emerging infectious diseases originate in wildlife reservoirs, yet the ecological and epidemiological processes that drive their emergence are often poorly characterised. Surveillance systems are typically fragmented across taxa, institutions, and spatial scales, resulting in data that are heterogeneous, biased, and incomplete. Wildlife surveillance, in particular, is constrained by logistical limitations and uneven sampling intensity, producing incomplete records that under-represent both the diversity and distribution of infected hosts \cite{schwind2014capacity,KUHN2024100929,deCock2024}. These gaps impede the ability to detect early emergence signals, quantify interspecies transmission risk, or forecast the spatial spread of novel pathogens.

Modelling provides a critical means of bridging these surveillance gaps. Mathematical and computational models, such as compartmental SEIR (Susceptible–Exposed–Infectious–Recovered) formulations, agent-based and network models, have been widely used to infer transmission processes, estimate key parameters, and simulate outbreak trajectories in both human and animal systems. However, zoonotic diseases pose unique challenges for such frameworks. Reservoir species often have poorly defined population structures, contact networks are shaped by dynamic ecological and behavioural factors, and pathogen detection is constrained by temporal and spatial sampling bias \cite{allen2012mathematical, White2018,roberts2021challenges}. Classical SEIR models assume homogenous mixing, while agent-based models often rely on static or incomplete representations of ecological interactions. Both approaches struggle to incorporate genomic data, environmental co-variates, and uncertainty arising from under-sampling.

Recent advances in graph-based and hybrid modelling offer new opportunities to integrate heterogeneous data sources into unified frameworks. Graph-based models can represent hosts, environments, and genetic relationships as interconnected nodes and edges, enabling multi-layer analyses that capture ecological complexity while retaining epidemiological interpretability. Such approaches are particularly well-suited to wildlife-borne zoonoses, where transmission pathways are shaped by spatial overlap, ecological affinity, and evolutionary proximity rather than direct observation.

To demonstrate this integrative approach, we focus on highly pathogenic avian influenza (HPAI) A/H5 as a representative case study. Avian influenza exemplifies many of the surveillance and modelling challenges common to wildlife zoonoses: broad host diversity, complex migratory behaviour, environmental persistence, and frequent cross-species transmission. The emergence and sustained spread of HPAI A/H5 clade 2.3.4.4b have reshaped the global landscape of avian and zoonotic disease risk. Once confined to poultry outbreaks and episodic wild bird mortality, these viruses are now established in migratory bird populations, with evidence of year-round enzootic circulation and transcontinental dispersal \cite{xie2023episodic,kandeil2023rapid}. In the United States, unprecedented waves of A/H5 outbreaks have swept through wild bird communities since late 2021, resulting in mass mortality events, ecological disruption, and continued spillover into domestic poultry and mammals \cite{caliendo2022transatlantic,kandeil2023rapid, peacock2025global,elsmo2023highly}. Over 14,000 wild bird detections have been reported across all major flyways, involving more than 150 species and marking the largest recorded HPAI epizootic in North America.

Despite extensive outbreak reporting, estimating the true prevalence and distribution of A/H5 infection in wild birds remains difficult. Surveillance largely depends on opportunistic carcass sampling and targeted testing of high-risk species, yielding datasets that are spatially biased and taxonomically sparse \cite{munster2007spatial,Hoye2010}. Consequently, gaps remain in understanding how A/H5 circulates across species, locations, and seasons. Addressing these gaps requires models capable of fusing ecological, genomic, and epidemiological information into coherent, interpretable structures.

In this study, we introduce \textsc{zooNet}, a graph-based epidemiological framework that integrates ecological transmission modelling, metadata-driven genetic imputation, and bi-layer heterogeneous graph learning to estimate pathogen prevalence and transmission potential in wildlife populations. While demonstrated here using A/H5 avian influenza in the United States during 2022, the framework generalises to other zoonotic systems, offering a scalable and data-efficient means to infer transmission dynamics in settings with incomplete surveillance and sparse genomic data.

\section{Overview of \textsc{zooNet}}\label{sec:overview}

\textsc{zooNet} is a modular, graph-based framework that integrates mechanistic transmission simulation, metadata-driven genetic distance imputation, and spatiotemporal graph learning within a unified representation of wildlife outbreak dynamics. Sparse surveillance detections are augmented using ecologically structured SEI simulations, and both observed and inferred cases are represented as nodes connected through spatial, temporal, host, and genetic relationships.These relationships are organised within a bi-layer heterogeneous graph in which outbreak events and administrative regions form interconnected layers. Spatial proximity, temporal adjacency, host similarity, and genetic relatedness are encoded as distinct edge types, with graph structure evolving across successive time steps.

\begin{figure}[H]
    \centering
    \includegraphics[width=1\linewidth]{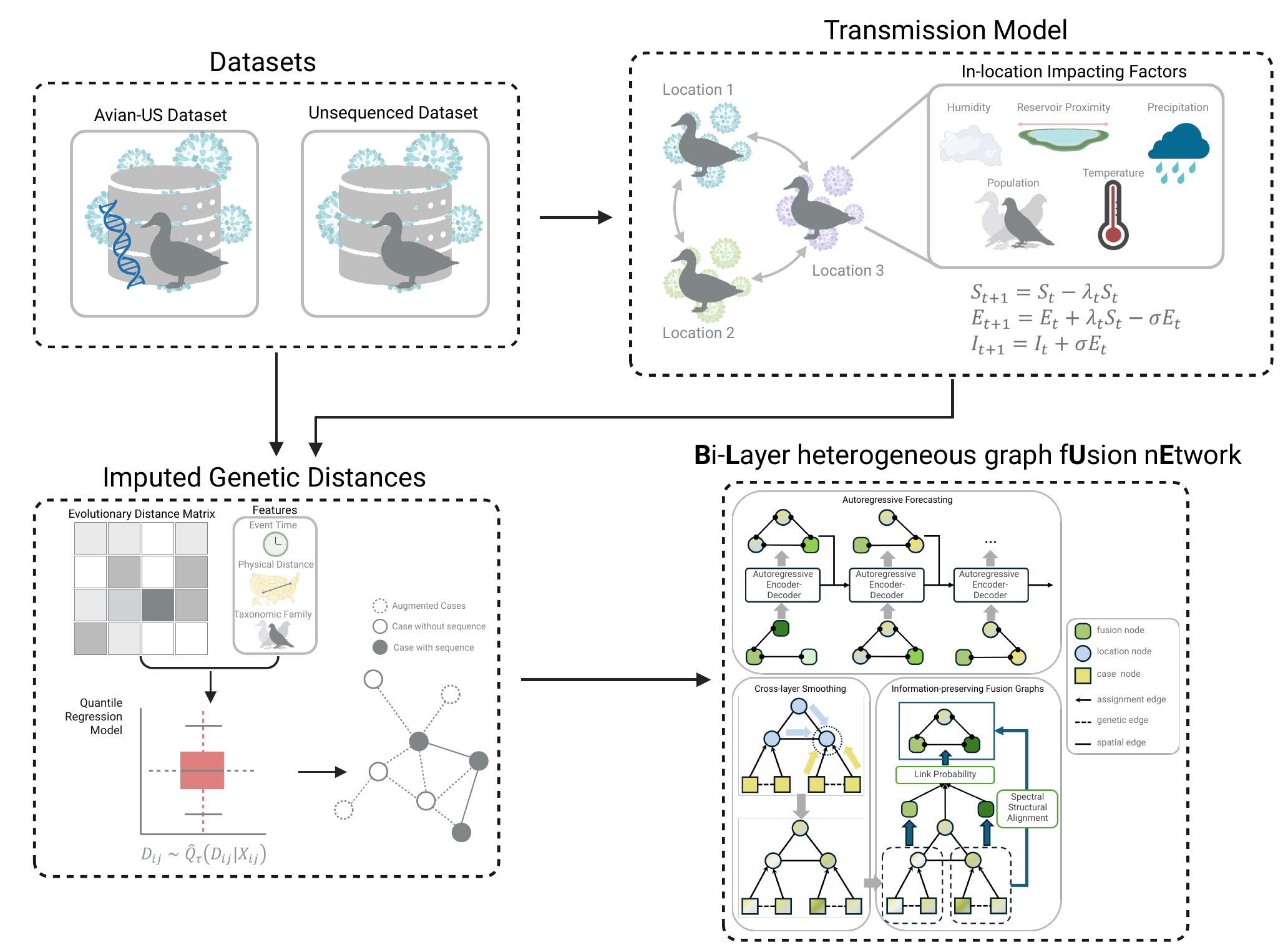}
    \caption{\textsc{zooNet} integrates ecological simulation, genetic distance imputation, and multilayer graph learning to forecast the spread of avian influenza A/H5 in wild birds. SEI-based simulation generates synthetic infections using environmental and host factors. Genetic distances are imputed using a metadata-driven quantile regression model. Cases are represented as nodes in a multilayer graph, linked by spatial, temporal, and genetic edges. Graph fusion and cross-layer smoothing are applied to unify these relationships, and temporal dynamics are captured through an autoregressive graph encoder–decoder.}
    \label{fig:framework}
\end{figure}

\section{Emergent Flyway Patterns of Transmission}
We evaluated \textsc{zooNet} separately for each US avian flyway using county-level case forecasts aggregated across all cross-validation folds. 
Overall performance varied by flyway, with the lowest mean squared error (MSE) observed in the Atlantic Flyway ($49.85 \pm 10.48$) and the highest in the Pacific Flyway ($155.81 \pm 48.91$). 
Detection performance, as measured by F1-score, ranged from $0.0565 \pm 0.0233$ in the Central Flyway to $0.1383 \pm 0.0725$ in the Pacific Flyway, reflecting differences in the precision--recall balance across regions. 
Pearson correlation measures the linear correlation between predicted and observed infection trends across spatial and temporal dimensions, and Spearman correlation identifies the ranking correlations between them.
Specifically, Pearson correlations between predicted and observed case counts were highest in the Pacific Flyway ($0.2173 \pm 0.0644$) and lowest in the Central Flyway ($0.0950 \pm 0.0280$). 
Regression error metrics indicated the smallest mean absolute error (MAE) in the Atlantic Flyway ($21.49 \pm 2.34$) and the largest in the Central Flyway ($29.86 \pm 3.86$). 
Detection recall was generally high across flyways (all above $0.58$), although precision remained low (all below $0.10$), indicating that the model was more effective at identifying true positive counties than at avoiding false positives. 
The infection rate predictions were highest in the Pacific Flyway ($0.0487 \pm 0.0110$), consistent with higher predicted case magnitudes in that region. 
These results suggest that while \textsc{zooNet} captures broad spatial patterns of outbreak occurrence, performance varies across flyways, with relatively stronger predictive accuracy in the Atlantic and Pacific regions.

\begin{table}[h]
\centering
\caption{Average \textsc{zooNet} cross-fold performance across for each US avian flyway. Values are mean $\pm$ standard deviation.}
\begin{tabular}{lcccc}
\toprule
\textbf{Metric} & \textbf{Atlantic} & \textbf{Central} & \textbf{Mississippi} & \textbf{Pacific} \\
\midrule
MSE & 49.8460 $\pm$ 10.4789 & 55.8994 $\pm$ 7.4182 & 55.1487 $\pm$ 18.3499 & 155.8075 $\pm$ 48.9092 \\
RMSE & 6.9205 $\pm$ 0.6361 & 7.4164 $\pm$ 0.5182 & 7.2659 $\pm$ 1.1603 & 12.2710 $\pm$ 2.0619 \\
MAE & 1.9422 $\pm$ 0.6242 & 2.1595 $\pm$ 0.7011 & 1.6345 $\pm$ 0.8504 & 3.3631 $\pm$ 0.9143 \\
F1 & 0.0854 $\pm$ 0.0400 & 0.0565 $\pm$ 0.0233 & 0.0815 $\pm$ 0.0325 & 0.1383 $\pm$ 0.0725 \\
\bottomrule
\end{tabular}
\label{tab:zooNet_flyway}
\end{table}

Across flyways, the ablation experiments show that the full zooNet model performs competitively relative to both the BLUE baseline and its intermediate variants, such as the augmented cases through the SEI model and genetic distance imputation. In the Central flyway, zooNet reduces MSE from 63.8 ± 14.4 to 55.9 ± 7.4, and in the Mississippi flyway from 53.7 ± 35.9 to 55.1 ± 18.3 with corresponding gains in RMSE and MAE. zooNet also achieves the highest F1 scores in the Pacific and Mississippi flyways at 0.138 ± 0.073 and 0.082 ± 0.033 respectively, exceeding all other variants. The augmented-case and genetic-distance variants show mixed effects, with improvements in some flyways and degradations in others, but none consistently outperform zooNet. 

\begin{figure}[H]
    \centering
    \includegraphics[width=1\linewidth]{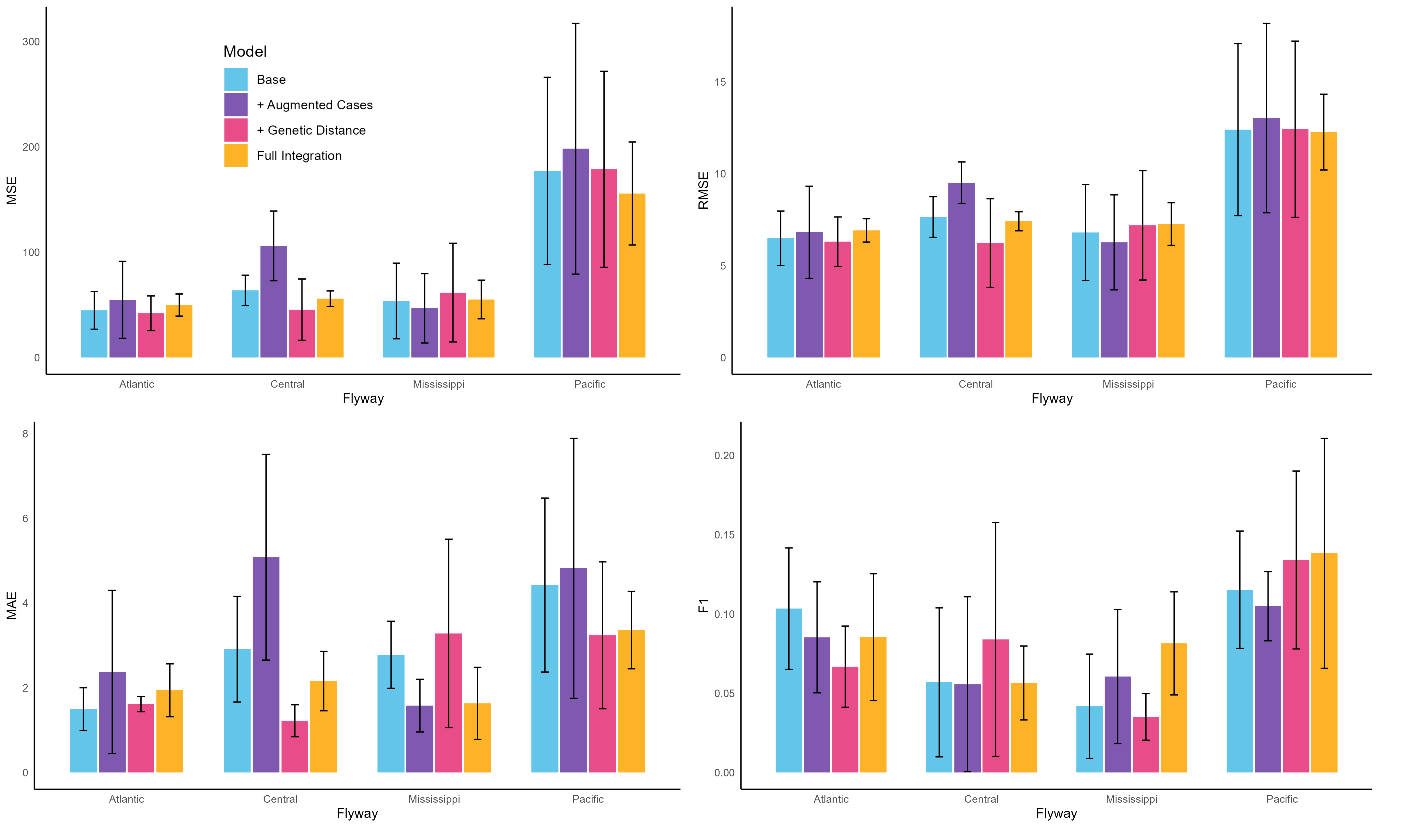}
    \caption{Effect of module removal on prediction error and
standard deviation across flyways.}
    \label{fig:placeholder}
\end{figure}

\section{Spatiotemporal Variability in Prediction}

\begin{figure}[H]
    \centering
    \includegraphics[width=1\linewidth]{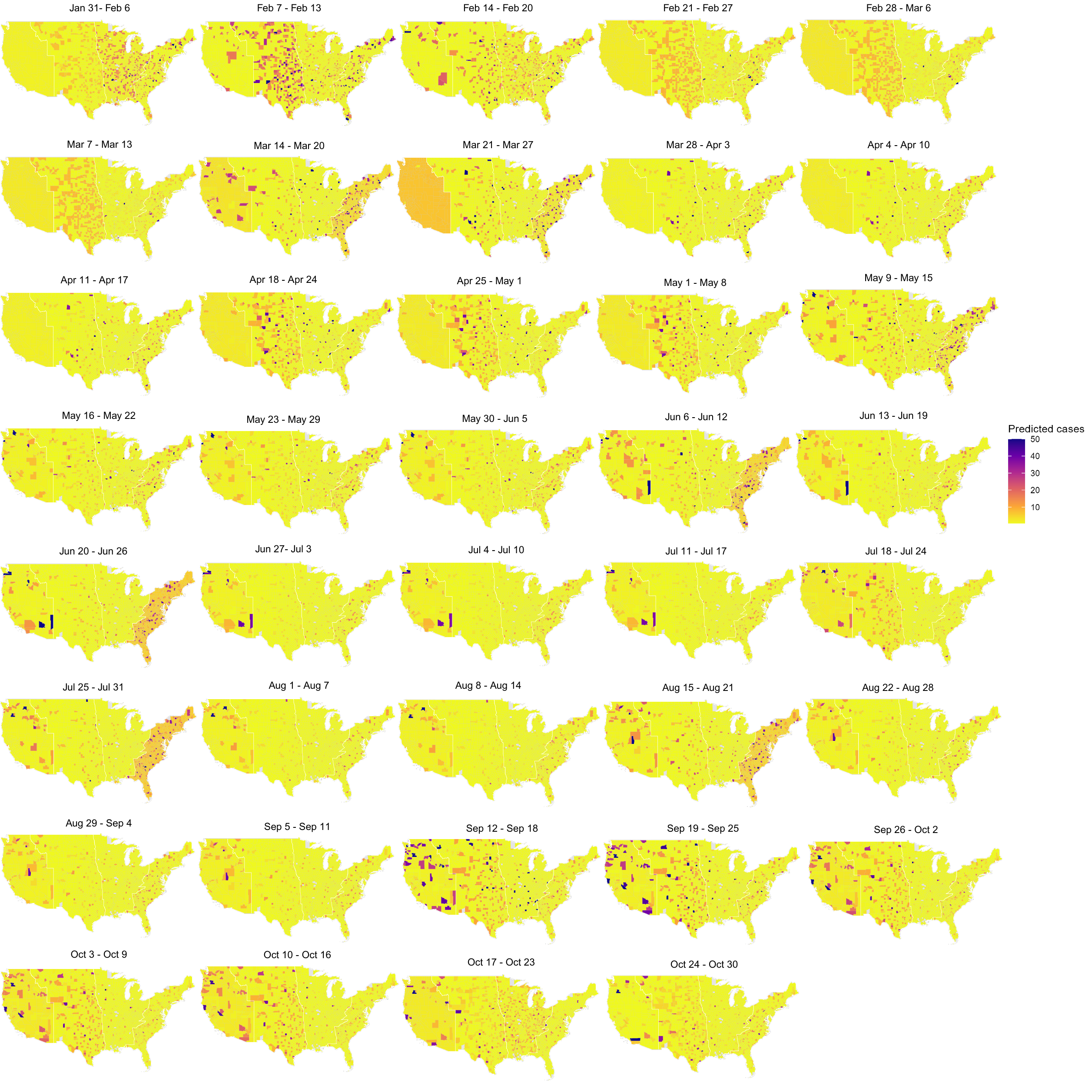}
    \caption{County-level case forecasts for the United States, stratified by avian flyway, simulated by \textsc{zooNet}, covering the weeks 31 January - 30 October 2022.}
    \label{fig:prediction}
\end{figure}

Across the prediction period (31~January–30~October), \textsc{zooNet} generated county-level forecasts that varied both spatially and temporally [Figure~\ref{fig:prediction}]. Predicted risk was generally concentrated along major migratory corridors, while large areas of the western United States maintained consistently low values throughout the season. Weeks with no new detections produced visually uniform maps in low-incidence regions.

Localised increases in predicted case counts often preceded reported outbreaks, producing transient spikes in mean predictions and local standard deviations. As an example of temporal variation at the county level, in New London County, Connecticut, the model predicted the onset of an outbreak beginning in the week starting 25~April~2022 (ISO week~17), before additional confirmed cases were reported, with sustained non-zero predictions in subsequent weeks. Predicted values continued through late September, culminating in the week starting 26~September~2022 (ISO week~39), when two new cases were observed.
Across flyways, counties that subsequently reported cases tended to have higher predicted values than those that did not (Pearson $r=0.10$–$0.22$, Spearman $\rho=0.09$–$0.14$), indicating a signal above baseline risk.

Temporal patterns differed between flyways. In the Atlantic Flyway, mean predicted cases were modest early in the season, peaking in weeks with observed detections (mean $\sim 3$–4 cases) before gradually declining. The Central and Mississippi Flyways showed sharper week-to-week fluctuations, reflecting both outbreak dynamics and intermittent surveillance detections (peak mean cases $\sim 4$–5). Pacific Flyway predictions remained relatively stable, with minor increases in weeks corresponding to observed events (mean cases $\sim 1$–2).

Across the full season, predicted distributions distinguished low-risk regions from localised increases in case counts, reflecting spatial heterogeneity and temporal variation in predicted risk.

\section{Migration Bottlenecks as Amplifiers of Viral Spread}

We estimated the overall prevalence of A/H5 avian influenza in wild birds to be 0.01 per 10,000. The overall prevalence varied seasonally, with the lowest values observed in winter (February, 0.004 per 10,000) when birds were largely concentrated in southern wintering areas. The prevalence increased through spring migration, peaking in May at 0.010 per 10,000, corresponding to the arrival of migratory waterfowl and other species on northern breeding grounds where high local densities favour viral transmission. During the summer months (June–August), prevalence remained moderate (0.006–0.007 per 10,000) as birds were widely dispersed across breeding territories, diluting transmission risk. In the fall (September–October), prevalence again rose slightly (0.005–0.005 per 10,000) as populations aggregated at staging sites during southward migration along major flyways, reflecting increased contact rates among migrating birds. These seasonal patterns of prevalence are closely aligned with the known migration and aggregation behaviours of wild bird populations in North America.

\begin{figure}[H]
    \centering
    \includegraphics[width=1\linewidth]{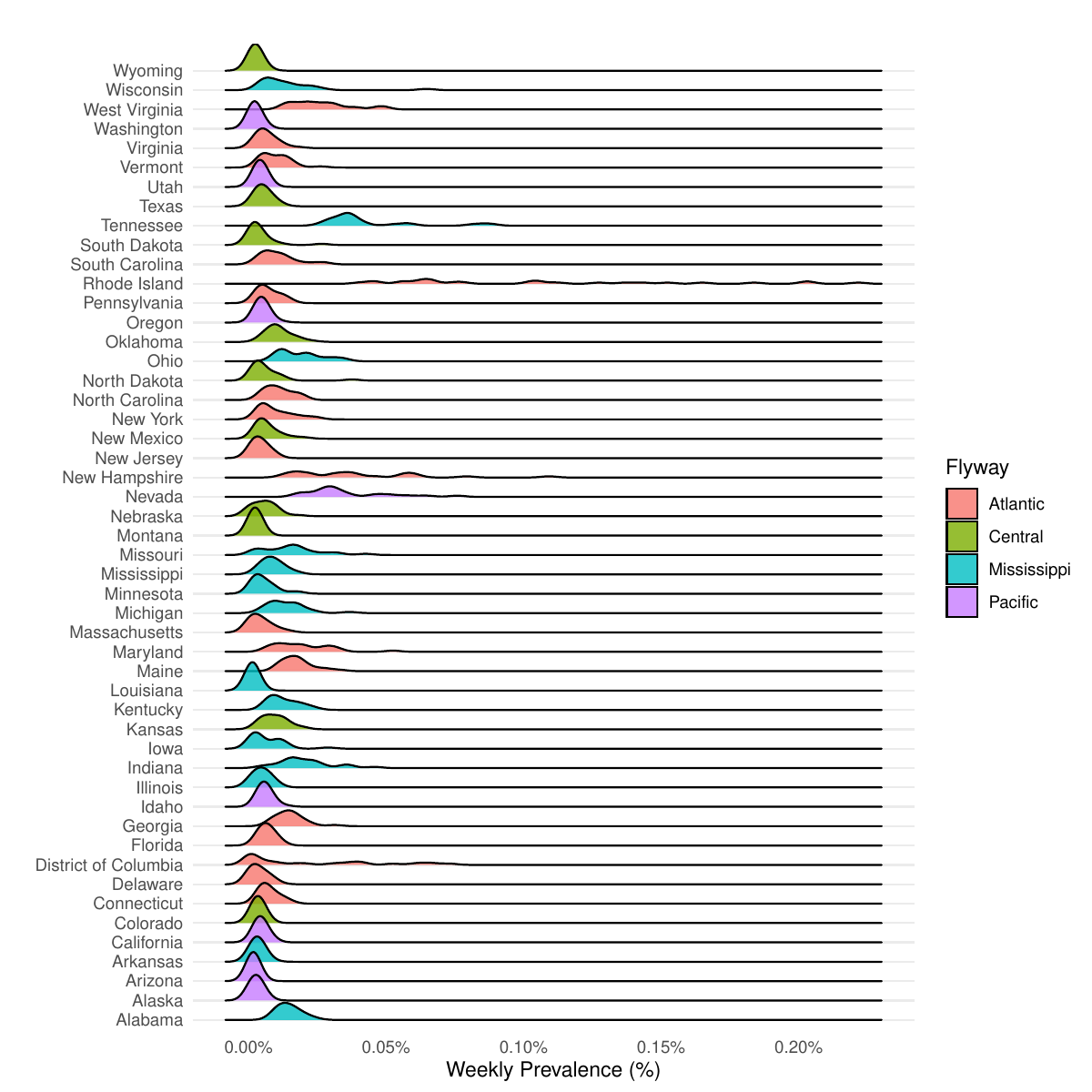}
    \caption{Weekly simulated prevalence, shown as percentage per 10,000, of A/H5 avian influenza in wild birds across North American states covering the weeks 31 January - 30 October 2022.}
    \label{fig:prevalence}
\end{figure}

Prevalence of A/H5 avian influenza varied markedly across individual states in North America (\ref{fig:prevalence}). The highest overall prevalence was observed in Virginia and Tennessee, with estimated mean values of 0.018 and 0.017 per 10,000, respectively, reflecting dense congregations of migratory waterfowl and active surveillance efforts. Tennessee also exhibited some of the most extreme weekly peaks, reaching 6.27 per 10,000 in late January and sustaining elevated prevalence into late spring (5.25–5.33 in May–August). Northern states such as Maine and New Hampshire consistently showed elevated prevalence across multiple time points, with Maine exhibiting early peaks in February (1.48–1.73 per 10,000) and late-season increases in September–October (up to 3.12), while New Hampshire displayed recurrent spikes in February–April (up to 4.61), indicating persistent viral circulation in these regions.

In contrast, western states including California, Washington, and Oregon exhibited lower prevalence, ranging from 0.004 to 0.006 per 10,000, consistent with more dispersed bird populations and reduced viral transmission. Smaller states and districts occasionally showed disproportionately high prevalence relative to the number of samples tested, reflecting localised outbreaks. Rhode Island, for example, displayed sharp spikes in February (3.29 per 10,000) and April (1.67), while the District of Columbia peaked in March–June at 2.42–2.62. Moderate prevalence was observed in northeastern states such as New York and Pennsylvania (0.012–0.013 per 10,000), where migratory stopover sites supported intermediate bird densities. Missouri showed late-season increases exceeding 4.4 per 10,000 in September, reflecting sustained outbreak activity in the Mississippi and Central Flyways. Seasonal trends were generally aligned across states, with prevalence peaking during spring migration (March–May) and again during fall migration (September–October), though the magnitude of these peaks was amplified in states with major staging areas.

At the county level, our simulations revealed pronounced spatial and temporal heterogeneity in A/H5 prevalence among wild birds. In Virginia, Staunton, Salem, and Bristol counties exhibited early winter peaks in January–February, reaching 12.4–19.2 per 10,000, reflecting congregations of waterfowl in southern wintering areas. Elevated prevalence persisted into spring, with Martinsville, Covington, Falls Church, Charlottesville, and Arlington showing sustained peaks of 6.1–14.0 per 10,000 in March–May, coinciding with northward migration and the arrival of breeding populations. Across the southeastern United States, extreme late-winter peaks were observed in Worth County, Missouri (77.0 per 10,000 in February) and Moore County, Tennessee (65.2 in January), while Quitman County, Georgia experienced its highest prevalence in March (36.0), highlighting the influence of migration bottlenecks and high-density staging habitats. Northern and western counties also displayed temporally distinct peaks, including Hancock and Unicoi counties in Tennessee (March–April, 10–22 per 10,000), Skagway in Alaska (May–June, 8–15), and Gilpin County, Colorado (July–August, 7–12), consistent with the seasonal timing of migratory stopovers and local aggregation.

\section{Discussion}\label{sec:Discussion}

Our results show that \textsc{zooNet} captures both the spatial and temporal complexity of highly pathogenic avian influenza (HPAI) A/H5 in North American wild birds while demonstrating how ecological, environmental, and data-driven processes that shape transmission in this system are broadly applicable across zoonotic diseases. Seasonal peaks in modelled prevalence coincided with spring and fall migration, reflecting the aggregation of migratory hosts at high-density staging sites where contact rates increase and interspecies mixing occurs \cite{roberts2021challenges, han2016global}. Conversely, summer dispersal across breeding territories diluted transmission, illustrating how spatiotemporal fluctuations in host density govern transmission potential across ecological contexts.

At regional scales, \textsc{zooNet} revealed that local ecological features, such as migratory bottlenecks and habitat connectivity, can drive transient yet intense infection clusters. These findings parallel dynamics observed in other multi-host zoonoses, where environmental suitability and host movement jointly modulate transmission opportunities \cite{gibb2020zoonotic, layan2021mathematical, allen2012mathematical}. Counties along major flyways, such as Tennessee and Missouri, exhibited pronounced but short-lived peaks, suggesting that spatially explicit ecological structure, rather than random stochasticity, explains much of the observed heterogeneity in outbreak intensity. This reinforces a key principle in zoonotic epidemiology: that ecological context, host behaviour, and landscape connectivity interact to determine where and when spillover and amplification are most likely to occur \cite{KUHN2024100929}.

Methodologically, the integration of SEI-based ecological simulation, evolutionary distance imputation, and graph-based model enables \textsc{zooNet} to preserve outbreak connectivity even under sparse and uneven surveillance \cite{roberts2021challenges, rees2021transmission}. High recall across flyways indicates that ecological and inferred genetic relationships can compensate for incomplete sampling, revealing latent epidemiological structure that is often obscured in purely spatial or count-based models. This property is not unique to avian influenza; similar graph-based integration could be applied to fragmented data from zoonotic systems such as Rift Valley fever, leptospirosis, or bat-borne henipaviruses, where incomplete surveillance and heterogeneous reporting constrain inference \cite{xue2013hierarchical,guo2023innovative,barker2013data,royce2020mathematically}.

Flyway-level differences further illustrate how ecological and surveillance contexts jointly shape predictive performance. Stronger correlations in the Atlantic and Pacific flyways reflected areas with consistent reporting and well-characterised migration pathways, whereas the Central and Mississippi flyways showed lower predictability due to patchier surveillance and ecological variability. This gradient mirrors broader zoonotic systems in which data density and environmental heterogeneity jointly determine model reliability, highlighting that prediction uncertainty can itself be a meaningful indicator of system complexity.

Comparisons to ecological and statistical baselines highlight the advantage of multi-modal data fusion. Models limited to spatial proximity or case counts fail to capture persistence, genetic linkage, or amplification through transient ecological hotspots \cite{Signore2025, adiga2024high}. By contrast, \textsc{zooNet} integrates ecological realism with inferred genetic connectivity to interpolate risk across unobserved times and regions, producing forecasts that maintain both spatial continuity and biological plausibility. 

Nonetheless, several limitations temper these findings. Sparse and uneven surveillance introduces unavoidable uncertainty, particularly in western and central regions. Environmental co-variates, while informative, cannot fully resolve microhabitat heterogeneity, interspecific interactions, or stochastic movement behaviours that influence transmission. Genetic imputation, though powerful, may overlook rare or emerging viral lineages, potentially biasing inferred outbreak connectivity. Similarly, stochastic variation within the SEI simulation propagates through the graph, contributing to variability in predicted outbreak magnitude. Absolute case counts should therefore be interpreted cautiously, with relative prevalence patterns providing a more reliable signal for ecological inference and surveillance prioritization.

Despite these constraints, \textsc{zooNet} offers a principled framework for understanding pathogen dynamics under realistic ecological and data limitations. By explicitly linking host abundance, movement, environmental context, and inferred genetic relationships, the model moves beyond presence-only risk mapping toward a population-level understanding of disease circulation. Its predictive and interpretive capacity provides a foundation for proactive surveillance design, highlighting high-risk periods and locations, and offers a blueprint for extending graph-based, multi-modal inference to other wildlife pathogens. More broadly, our results illustrate that even with sparse and biased data, integrating ecological realism with network-based inference can reveal latent epidemiological structure and generate mechanistically grounded forecasts, a critical step for anticipating and mitigating outbreaks in complex wildlife systems.

In summary, \textsc{zooNet} demonstrates how surveillance data that are sparse, heterogeneous, and biased toward detections can be transformed into ecologically interpretable forecasts of HPAI circulation. The framework recovered broad spatiotemporal patterns of viral risk across North American flyways, with strongest performance in regions where detections were more consistent and ecological connectivity clearer, while still producing meaningful forecasts under data-poor conditions. By incorporating a prevalence layer, \textsc{zooNet} moves beyond presence-only risk mapping toward estimates that link surveillance detections to infection dynamics at the population level, offering a bridge between opportunistic monitoring and ecological inference. Quantitative gains over ecological and statistical baselines highlight the benefit of graph-based fusion for integrating multi-modal data, but perhaps more importantly, the results illustrate how such integration can provide useful signals even when surveillance is incomplete. In this way, \textsc{zooNet} offers a pragmatic step toward improving early warning and interpretation of wildlife disease data, while remaining constrained by the limitations of the available surveillance.

\section{Methods}\label{sec:Methods}

\subsection{Graph Construction and Layer Definition}\label{sec:construction-layerdef}

The \textsc{zooNet} Graph Neural Network framework represents outbreak dynamics using a bi-layer heterogeneous graph structure, adopting the architecture of BLUE \cite{du2025bluebilayerheterogeneousgraph} for representation learning and message passing. The model operates on a multilayer graph \( G = (V, E) \), where the node set \( V \) contains two distinct types: outbreak events and administrative regions. The edge set \( E \) encodes spatiotemporal, host-related, and genomic relationships between nodes.

The event layer of the graph comprises nodes with genomic importance features, representing both observed surveillance cases and synthetic infections generated from a mechanistic susceptible--exposed--infectious (SEI) model. SEI simulations incorporate external ecological and environmental information to generate outbreak trajectories, but the graph itself contains only outbreak nodes characterised by spatial coordinates, detection dates, and host species attributes. SEI-augmented nodes inherit attributes from their simulated trajectories and are embedded into the graph identically to observed cases.

Edges within the event layer capture pairwise relationships derived from multiple sources. Each edge \( e_{ij} \in E \) connecting nodes \( i \) and \( j \) is assigned a set of co-variates,
\[
e_{ij} = \left( d_{ij}^{\text{spatial}}, \Delta t_{ij}, s_{ij}^{\text{host}}, d_{ij}^{\text{genetic}} \right),
\]
where \( d_{ij}^{\text{spatial}} \) denotes geographic distance, \( \Delta t_{ij} \) represents temporal lag, \( s_{ij}^{\text{host}} \) quantifies host-related similarity, and \( d_{ij}^{\text{genetic}} \) corresponds to genetic divergence. Genetic distances are computed from viral sequences when available and imputed via a predictive model trained on outbreak-specific co-variates for unsequenced or simulated nodes. Edge weights are represented consistently across empirical, simulated, and imputed cases.

The region layer consists of administrative units, such as counties or flyways, with outbreak event numbers and abundance as region features. Regions are connected by edges that encode geographical adjacency or other forms of spatial proximity. Cross-layer edges link outbreak nodes to their corresponding regions, enabling multi-scale information exchange between outbreaks and broader spatial contexts.

\subsection{Cross-layer Smoothing and Graph Fusion}\label{subsec:smoothing-fusion}

To enable coherent learning across layers, \textsc{zooNet} incorporates a cross-layer smoothing mechanism inspired by Markov random fields. This mechanism allows information to propagate between the outbreak and region layers, ensuring that spatial context can influence outbreak-level predictions and vice versa. The smoothing process is formalised through graph Laplacian regularisation, which penalises discrepancies between node embeddings across layers.

In addition, \textsc{zooNet} implements an information-preserving graph fusion strategy to integrate ecological, genetic, and spatial relationships into a unified graph representation. This fusion process employs spectral regularisation to minimise structural distortions during the combination of heterogeneous graphs. By preserving key graph properties, such as node connectivity patterns and local outbreak topology, the model maintains fidelity to the original outbreak relationships while unifying diverse data types.

\subsection{Temporal Graph Sequence Modelling}\label{subsec:temporalgraphseq}

Beyond static modelling, \textsc{zooNet} captures the temporal dynamics of avian influenza outbreaks through an autoregressive graph sequence modelling framework. The model constructs a sequence of graphs \( \{G_t\} \) over successive time intervals, with each graph encoding the outbreak and region structure at a given time point.

To model temporal evolution, \textsc{zooNet} applies a graph neural network-based encoder-decoder architecture. The encoder aggregates information from past graph snapshots to learn dynamic node representations that capture both short-term and long-term outbreak dependencies. The decoder then predicts future graph states, enabling the model to forecast the progression of outbreaks across space and time. 

\subsection{Augmented case simulation via ecologically structured SEI modelling}\label{subsec:augmentedSEI}

To estimate short-range transmission potential of avian influenza A/H5 in settings where direct transmission linkages are unobservable, we developed a Susceptible–Exposed–Infectious (SEI) simulation framework that incorporates local ecological and environmental conditions.The transmission coefficient $\beta_t$ was modelled as a dynamic function of four ecologically relevant variables: temperature, relative humidity, precipitation, and proximity to water reservoirs \cite{biswas2014modeling,lowen2007influenza,si2013different,islam2023potential}. These variables influence both viral viability and host contact patterns, and their effects were incorporated through empirically supported response functions.

Temperature, relative humidity, and precipitation were modelled using Gaussian response functions to capture non-linear relationships where transmission peaks at intermediate conditions and declines under extremes \cite{lowen2007influenza,peci2019effects,deyle2016global}.Proximity to aquatic habitats, which are critical congregation sites for reservoir hosts, were modelled via an exponential decay function to represent the declining likelihood of transmission with increasing distance from water \cite{si2013different,gilbert2007avian}:

\[
f_W(W_r) = \exp(-W_r / \theta),
\]

where $W_r$ denotes the distance to the nearest reservoir and $\theta$ is a fitted decay parameter. The Gaussian response functions for the meteorological variables take the form:
\[
f_X(X_r) = \exp\left(-\frac{(X_r - X_{\text{peak}})^2}{2 \sigma_X^2}\right),
\]
where $X_r$ represents the local environmental variable (temperature $T_r$, humidity $H_r$, or precipitation $P_r$), $X_{\text{peak}}$ is the optimal value for transmission, and $\sigma_X$ sets the tolerance range.

Environmental response parameters were modulated by hydrological context. Sites near water bodies had distinct optima and tolerances, reflecting differences in host behaviour and viral persistence between wetland and terrestrial settings \cite{hill2022ecological,si2013different}. Specifically, temperature optima were set lower near water ($10^\circ\mathrm{C}$) and broader ($\sigma_T = 8$) than at inland sites ($15^\circ\mathrm{C}$, $\sigma_T = 6$); analogous adjustments were applied for humidity and precipitation (Supplementary Information).

The total transmission coefficient at each time step was computed as:
\[
\beta_t = \beta_0 \cdot f_W(W_r) \cdot f_T(T_r) \cdot f_H(H_r) \cdot f_P(P_r),
\]
where $\beta_0$ is the baseline transmission rate.

Each outbreak was treated as a discrete transmission initiation event, annotated with spatial coordinates, detection date, and local bird abundance derived from eBird spatiotemporal species distribution models. Total abundance $N$ was extracted from remotely sensed raster layers, resolving species-level presence within each outbreak vicinity. The SEI model was initialised with $S_0 = N - 1$, $E_0 = 1$, and $I_0 = 0$, assuming the detected outbreak corresponds to a single exposed index case within the host population.

Simulations were run over a fixed 10-day window using a discrete-time Euler approximation:
\[
S_{t+1} = S_t - \lambda_t S_t, \quad E_{t+1} = E_t + \lambda_t S_t - \sigma E_t, \quad I_{t+1} = I_t + \sigma E_t,
\]
where $\lambda_t = \frac{\beta_t I_t}{N}$ is the force of infection and $\sigma = 0.4$ represents the rate of progression from exposed to infectious compartments, consistent with observed A/H5 latent periods of 2–3 days in experimental studies. Given the high virulence of circulating A/H5 clade 2.3.4.4b and surveillance bias toward carcasses, we truncated onward transmission at the exposed/infectious boundary, without explicitly modelling recovery or mortality.

At each simulation step, $\beta_t$ was recomputed using daily environmental data, allowing transmission potential to vary in response to meteorological and ecological fluctuations. This produces site-specific infection trajectories $\{S_t, E_t, I_t\}$, representing the expected local transmission envelope under prevailing conditions at the time of detection.

\subsection{Genetic distance imputation for unsequenced and augmented cases}\label{subsec:geneticdistance}

To address the challenge of missing genomic data, we implemented a biologically informed imputation framework that estimates evolutionary distances for all outbreaks lacking sequenced isolates, including both reported and augmented infections \cite{stone2025probabilisticframeworkimputinggenetic}. Let \( D_{ij} \) denote the evolutionary distance between case \( i \) and case \( j \). For case pairs where one or both nodes lack viral sequence data, \( D_{ij} \) is estimated using a quantile regression model \( \widehat{Q}_{\tau}(D_{ij} \mid \mathbf{X}_{ij}) \), where \( \mathbf{X}_{ij} \) is a vector of pairwise co-variates, and \( \tau \in \{0.05, 0.50, 0.95\} \) is the quantile of interest.

The Kimura two-parameter (K80) model was selected as the evolutionary distance metric due to its suitability for short-timescale divergence typical of seasonal or epizootic avian influenza dynamics. Unlike the simpler Jukes-Cantor (JC69) model, K80 accounts for transition/transversion bias, a feature of influenza virus evolution associated with mutational mechanisms and host-specific editing processes. While more complex substitution models such as TN93 or GTR offer additional flexibility, they require more parameters and are less stable when applied to sparse or incomplete alignments, which are common in avian influenza surveillance data. 

The co-variate set \( \mathbf{X}_{ij} \) includes host family, spatial separation, temporal lag (\( \Delta t \)), administrative proximity, and ecological similarity. The model was trained on a large set of outbreak pairs with known sequences using heteroskedastic gradient boosting, yielding robust interval estimates across the full spectrum of genetic divergence.

For observed outbreaks without sequences, imputed distances to all other nodes with sequences were computed using the above model, preserving the outbreak’s genomic connectivity within the graph. For augmented cases, metadata were inherited directly from the parent SEI simulation, including location, date, and host species. The imputation model was then applied to generate a full set of pairwise distance distributions \( \widehat{Q}_{\tau}(D_{ij}) \), and the resulting estimates were retained as quantile-valued edge weights in the outbreak-level graph.

\subsection{Experimental Setup}
We evaluate our method on the self-collected \textsc{zooNet} dataset (3,227 U.S. counties with aligned genetic, geographic, and ecological modalities). In each experiment, the model conditions on a history window of $T=4$ weeks and forecasts the subsequent $H=4$ weeks. Evaluation metrics (RMSE, MAE, PCC, F1) are averaged across the $H$ steps. We adopt 5-fold cross-validation (80\% for training and 20\% for testing on each fold) for all models using a fixed random seed for reproducibility, and report fold-averaged results.

For the genetic importance feature of each observed node in the event layer, we average all the non-zero genetic distances between the node and its neighbors and subtract this distance by 1 to represent genetic similarity. 
For fairness, all baselines use a unified embedding dimension $d=8$ and a dropout rate of $0.3$. For \textsc{zooNet} specifically, we employ $K=3$ cross-layer smoothing iterations, and set the number of layers in the autoregressive temporal graph sequence modelling to $L=2$.

We optimize the composite objective $\mathcal{L}_{\mathrm{tot}}
     \;=\;
     \underbrace{\frac{1}{H}\sum_{h=T+1}^{T+H}\sum_{i=1}^{N}
       \bigl(\hat y_{i,h}-y_{i,h}\bigr)^{2}}_{\mathcal{L}_{\mathrm{pred}}}
     \;+\;
     \lambda_{\mathrm{1}}\,\cdot
     \underbrace{\bigl\|\,\mathbf{L}_{hetero} - \mathbf{L}_f \bigr\|_{F}^{2}}_{\mathcal{L}_{\mathrm{spec}}}
     \;+\;
     \lambda_{2}\,
     \underbrace{\sum_{w\in\Theta}\|w\|_{2}^{2}}_{\mathcal{L}_{r}}$
where $\mathcal{L}_{\mathrm{pred}}$ is multi-step MSE over the forecasting horizon of the temporal graph sequence modelling (Sec. \ref{subsec:temporalgraphseq}), the spectral term $\mathcal{L}_{\mathrm{spec}}$ aligns the Laplacian of the learned fusion graph with that of the original heterogeneous graph to achieve information-preserving fusion (Sec. \ref{subsec:smoothing-fusion}), and the last term $\mathcal{L}_{r}$ is standard weight decay. 
The spectral alignment weight is set as $\lambda_1=0.2$ based on empirical results.
Models are trained for 100 epochs with learning rate $1e-5$ and weight regularization $\lambda_2=5e-4$, with an early stopping mechanism activated if no improvement on the training set is observed for 10 consecutive epochs. 

\subsection{Ablation Study}
To quantify the contribution of each modelling component, we conducted ablation experiments in which the SEI-augmented cases and the genetic-distance imputation module were removed individually while keeping all other architecture, hyperparameters, and training procedures fixed. For each flyway, models were trained and evaluated under the same 5-fold cross-validation scheme described above, using identical history and forecasting windows. The “Base” model corresponds to the BLUE framework applied to the empirical outbreak graph without augmented nodes or imputed genetic edges. The “+ Augmented Cases” variant adds SEI-generated nodes but retains only observed genetic distances, whereas the “+ Genetic Distance” variant supplies imputed genetic edges for all unsequenced cases but excludes SEI-augmented nodes. Performance metrics (MSE, RMSE, MAE, F1) were averaged over folds for all variants to isolate the marginal value of each module.
\subsection{\textsc{zooNet} Dataset}


The AvianUS dataset comprises 12,333 avian influenza A/H5 cases and more than 8,000 associated hemagglutinin (HA) sequences reported in the United States between 2021 and 2024. For the \textsc{zooNet} subset we restricted to wild bird cases from 2021 to 2022, yielding 7,362 cases and 2,143 sequences. Wild birds were prioritized because they act as reservoirs and long-distance dispersal hosts for highly pathogenic avian influenza, providing insight into natural transmission pathways across landscapes. In contrast, poultry outbreaks are typically localized and strongly influenced by farm management, and after 2024 national surveillance shifted toward dairy cattle, limiting continuity in wildlife monitoring. The 2021–2022 focus therefore captures the critical early phase of clade 2.3.4.4b circulation in North American wild birds.

Case records were obtained from federal surveillance systems and standardised at the county level. The host descriptors were harmonised using the International Ornithological Congress taxonomy, which resolved ambiguous or under-specified entries to species-level identifiers. Viral HA sequences were retrieved from public repositories and filtered to retain only wild bird hosts. A supervised record linkage model, trained on labelled match and non-match examples, associated sequences with case records using taxonomic agreement, spatial proximity, and temporal overlap within 14 days.The model used gradient-boosted decision trees to score candidate pairs, and high-confidence matches were retained for analysis. Wild bird abundance data were derived from the eBird Status and Trends product, aggregated to county-week resolution, and temporally aligned with case records \cite{eBird_Weekly_Abundance}.

\section*{Acknowledgements}
This research is supported by the Australian Commonwealth Scientific and Industrial Research Organisation (CSIRO) and the the United States National Science Foundation under Grants No. 2302968, No. 2302969, and No. 2302970, titled "Collaborative Research: NSF-CSIRO: HCC: Small: Understanding Bias in AI Models for the Prediction of Infectious Disease Spread". This research is conducted by the ARC Centre of Excellence for Automated Decision-Making and Society (No.CE200100005) funded by the Australian Government through the Australian Research Council.We acknowledge the utilization of computational resources from the Katana High Performance Computing (HPC) cluster, which is supported by the Faculty of Engineering, UNSW Sydney. We also acknowledge the National Computational Infrastructure (NCI) for providing access to the Gadi
supercomputer.

\section*{Data and Code Availability}
The source code for \zooNet\ is available on GitHub (\url{https://github.com/cruiseresearchgroup/zooNet}).

\bibliographystyle{plainnat} 
\bibliography{sn-bibliography}

\begin{appendices}
\end{appendices}

\end{document}